\documentclass[conference]{IEEEtran}
\IEEEoverridecommandlockouts
\usepackage{cite}
\usepackage{amsmath,amssymb,amsfonts}
\usepackage{algorithmic}
\usepackage{graphicx}
\usepackage{textcomp}
\usepackage{xcolor}
\usepackage{eucal}
\usepackage{booktabs}
\usepackage{multirow, makecell}
\def\BibTeX{{\rm B\kern-.05em{\sc i\kern-.025em b}\kern-.08em
    T\kern-.1667em\lower.7ex\hbox{E}\kern-.125emX}}

\begin{document}

\title{Effect of Duration and Delay on the Identifiability of VR Motion
\thanks{
Funded through the Stanford Sozo Graduate Fellowship, 2022-2023, and NSF Award 1800922}
}

\author{\IEEEauthorblockN{Mark Roman Miller}
\IEEEauthorblockA{\textit{Computer Science} \\
\textit{Illinois Institute of Technology}\\
Chicago, USA \\
mmiller30@iit.edu}

\and
\IEEEauthorblockN{Vivek Nair}
\IEEEauthorblockA{\textit{Computer Science} \\
\textit{UC Berkeley}\\
Berkeley, USA \\
vcn@berkeley.edu}

\and
\IEEEauthorblockN{Eugy Han}
\IEEEauthorblockA{\textit{Communication} \\
\textit{Stanford University}\\
Stanford, USA \\
eugyoung@stanford.edu}

\and
\IEEEauthorblockN{Cyan DeVeaux}
\IEEEauthorblockA{\textit{Communication} \\
\textit{Stanford University}\\
Stanford, USA \\
cyanjd@stanford.edu}

\and
\IEEEauthorblockN{Christian Rack}
\IEEEauthorblockA{\textit{Human-Computer Interaction} \\
\textit{University of Würzburg}\\
W{\"u}rzburg, Germany \\
christian.rack@uni-wuerzburg.de}

\and
\IEEEauthorblockN{Rui Wang}
\IEEEauthorblockA{
\textit{Carnegie Mellon University}\\
Pittsburgh, USA \\
ruiwang3@andrew.cmu.edu}

\and
\IEEEauthorblockN{Brandon Huang}
\IEEEauthorblockA{
\textit{UC Berkeley}\\
Berkeley, USA \\
zhaobin@berkeley.edu}

\and
\IEEEauthorblockN{Marc Erich Latoschik}
\IEEEauthorblockA{\textit{Human-Computer Interaction} \\
\textit{University of Würzburg}\\
W{\"u}rzburg, Germany \\
marc.latoschik@uni-wuerzburg.de}

\and
\IEEEauthorblockN{James F. O'Brien}
\IEEEauthorblockA{\textit{Computer Science} \\
\textit{UC Berkeley}\\
Berkeley, USA \\
job@berkeley.edu}

\and
\IEEEauthorblockN{Jeremy N. Bailenson}
\IEEEauthorblockA{\textit{Communication} \\
\textit{Stanford University}\\
Stanford, USA \\
bailenso@stanford.edu}

}

\maketitle

\begin{abstract}
xvSocial virtual reality is an emerging medium of communication. In this medium, a user's avatar (virtual representation) is controlled by the tracked motion of the user's headset and hand controllers. This tracked motion is a rich data stream that can leak characteristics of the user or can be effectively matched to previously-identified data to identify a user. To better understand the boundaries of motion data identifiability, we investigate how varying training data duration and train-test delay affects the accuracy at which a machine learning model can correctly classify user motion in a supervised learning task simulating re-identification. The dataset we use has a unique combination of a large number of participants, long duration per session, large number of sessions, and a long time span over which sessions were conducted. We find that training data duration and train-test delay affect identifiability; that minimal train-test delay leads to very high accuracy; and that train-test delay should be controlled in future experiments.

\end{abstract}

\begin{IEEEkeywords}
virtual reality, identifiability, privacy, duration, delay
\end{IEEEkeywords}

\section{Introduction}
Recently, virtual reality (VR) has been increasing in popularity, including the use case of social VR. Social VR is a medium in which users, represented by virtual characters called avatars, interact in a shared virtual space. Instances of this include VRChat, Horizon Workrooms, RecRoom, and Gorilla Tag.
If this medium becomes a mainstay in the consumer space, it will be important
to discover, understand, and address the risks associated with its
use.
%It is clear there is a dramatic risk to privacy in this setting. 
%In order to render a shared virtual world on many different devices, the status of objects and people within that world need to be shared across all those devices. In contrast to some public spaces, where anything can \textit{potentially} be recorded, social VR public spaces, by nature of being virtual, are \textit{actually} always being recorded. 

One risk we focus on in this work is re-identification attacks enabled by the rich nonverbal behavior that VR captures, from which behavioral biometrics can be inferred \cite{Pfeuffer2019,Moore2021,Miller2020,Nair2022}.

Behavior changes over time, and so it is valuable to know to what degree the behavioral biometric from motion collected at one time can be applied to motion collected at another time. Understanding the long-term temporal stability of motion biometrics can help determine whether accessing a user's motion data is a temporary risk, like a password breach, or a long-term risk, like other biometrics such as fingerprints. Overall, this work provides several contributions to our understanding of motion as biometric:
\begin{itemize}
\item findings corroborating previous work \cite{Moore2021} that identifiability
is higher within a session than between separate sessions
\item results indicating the delay between training data and testing data
affects identifiability in the range from one to seven weeks (subsection \ref{subsec:Delay})
\item results indicating short samples taken over several sessions are more
identifying than longer samples in fewer sessions (subsection \ref{subsec:Duration})

\end{itemize}

\section{Related Work}

We describe the landscape of identification by motion, with a particular focus on identification over time.

In this review and throughout the paper, we interleave references to security-focused and privacy-focused literature. In both cases, someone (the authenticator or attacker) is identifying a user based upon the user's data, so there is a fundamental similarity of mechanism. However, the design considerations and social settings are different.

\subsection{Identification Using Motion Intended for Identification}

There is a fair amount of work on use of VR pose information as a behavioral biometric, but much of it investigates an entity (authenticator or attacker) who has access to more than just the motion data. 

One thread of work studies which combination of an action and a matching algorithm can produce an effective ``motion password" \cite{Olade2020,mathis_knowledge-driven_2020,Wang2021,liebers_using_2021}. These works presume an overt authentication method and cooperation from the user. Another thread of work explores covert, cooperative elicitation of a certain kind of motion using social interaction. For example, an attacker can elicit certain actions from a target by waving at another user in social VR \cite{Falk2021} or throwing a ball to a user and expecting the user to throw it back \cite{Miller2021,Miller2022,miller2020within}. A third thread elicits the user's cooperation through the design of a virtual world \cite{Nair2022}. 

All of these methods require an attacker to have more capabilities than simply access to the motion data, either by overt cooperation, covert cooperation through interaction and social norms, or manipulation of the environment. They all presume a relatively stronger attacker than we assume in this work, where we focus on motions that are natural for a task other than identification, yet can still be used to identify a user.

\subsection{Identification using Natural Motions}

The focus of this work is on the ability for motion data \textit{alone} to be the identifying factor. With 'motion' being such a broad category, how can an authenticator or attacker ensure the user performs the particular kind of motion necessary to identify them? Previous work takes one of three approaches to this question. The first is selecting reasonably common actions that a user would perform anyway, such as operating 3DUI elements \cite{Pfeuffer2019} or walking \cite{Shen2018}. Second, it is plausible to build a model that explicitly learns a representation of motion across different kinds of action, such as is done by Rack and collaborators \cite{schell_extensible_2023}.

The third approach, which is the approach we take in this work, is to dismiss the notion that one needs to select a certain kind of motion - any kind of motion will do. For example, there is watching 360-degree video \cite{Miller2020,sivasamy_vrcauth_2020,wierzbowski_behavioural_2022}, training in a surgery room simulator \cite{Moore2021}, or playing a VR game \cite{mustafa_unsure_2018} with Beat Saber being a common subject \cite{liebers_exploring_2023,nair_unique_2023}. In contrast to these works with a similar threat model, the present work examines the effect of time much more in-depth.

\subsection{Identification Over Time}

In this work, we focus specifically on identification over time. The
delay in time between a user's training data (\textit{i.e.}, enrollment) and
a user's testing data (\textit{i.e.}, query) seems to affect accuracy, with longer delays leading to worse accuracy. While few works have investigated this directly, it is possible to infer a trend based on a review of the literature. For example, a delay of 30 seconds
between sessions and data collected over the span of about an hour
found 98\% accuracy \cite{Wang2021}, no delay and a span of 10-15
minutes found 95\% accuracy \cite{Miller2020}, sessions recorded
on ``different days'' found 90\% accuracy \cite{Liebers2021},
and one week later found 42\% accuracy \cite{Moore2021}. In general, shorter delays seem to imply higher accuracies.

This effect
of time delay is explicitly studied by R. Miller and collaborators
\cite{Miller2022} by combining two sets of data collected up to 18
months apart. They find no effect of delay on short-scale separations
(within 24 hours) or medium-scale separations (comparing delays shorter
than 3 days and longer than 3 days in one analysis, and the same but
for 10 days in a second analysis). On long timescales, which in their
work goes from 7 to 18 months, there were changes in behavior and
a reduction in accuracy, which was not the case in the short- and
medium-term delays. However, the delays were not regularly spaced
and some varied widely in magnitude. While that work established that identifiability changed over time, it is still an open question
how identifiability changes over time.

%[Deep dive into multiple sessions, etc.]
To summarize the contrast to previous work, we focus on a common social
VR activity, specifically, group discussion. This focus presumes a weaker attacker that does not need to be trusted by the user, to be present with the user in the VR environment, or to be the designer of the environment the target is in. In the dataset, there are regular spacing of data
collection periods which supports a more systematic estimation of the rate at which identifiability decays. Finally, the data we have collected for analysis has a larger sample size than most, more
collected data than almost any other, and it was collected over a longer
duration than most.

\section{Methods}

\subsection{Threat Model}

We characterize our threat model on two dimensions. First, there is the question
of what data is available to the attacker. Using the taxonomies given by Nair and collaborators
\cite{Nair2022} and Garrido and collaborators \cite{garrido2023sok}, we focus on the \textit{unprivileged user}, who has access only to the data provided by other users of a hypothetical social VR application. Second, we also restrict the kind of influence the attacker has on the user, making attacks like designing a virtual world specifically for identification or contrived interactions with a user out-of-scope.

This threat actor is selected because it is the least privileged attacker, making the attack most widely available. Additionally, there are some cases
in which this mode of attack may be the only available to an attacker. Examples include large-scale surveillance
where individuals are not queried directly, re-identification attacks
where actions are stored for a period of time before being queried,
or any other situations in which the attacker does not have
any direct interaction with the target.

\subsection{Data}

The data used in this work comes from the Stanford Longitudinal Virtual Reality Classroom Dataset (SLVRClaD) \cite{Han2023}, which is available upon request from the original study's authors.
SLVRClaD consists of two periods of data collection
of classroom immersive VR. A total of 232 participants met in small groups ranging from
two to 12 and consented to have their verbal, nonverbal, and performance data
continually tracked during each course. The course included eight weekly sessions
that lasted about 30 minutes per session. The current paper utilizes
previously unreported data from the dataset, and focuses on identifiability
of this nonverbal pose data.

Weekly activities varied, but included both large and small group discussion as well as VR
building activities. Sessions were led by a researcher. See \cite{Han2023} for further details of activities.

The motion data collected consisted of the position and orientation of the participants' headsets and hand controllers in world-space coordinates at a nominal framerate of 90 frames per second. Of the original 232 participants, the data used in this study consisted of 183 participants who had at least 5 unique sessions (of 8 possible) and at least 2 hours of tracked motion data in total.

\subsection{Feature Engineering}

% The feature set we used consisted of 840 \textit{streams} that were subset and
% summarized in various ways. Some of these streams were defined in
% terms of body-space coordinates, which is described below.

Because the identifiable features of one's pose are often invariant to rotations
within the horizontal plane, we normalize this motion data using \textit{body-relative coordinates} \cite{schell_extensible_2023,rack_who_2023}.
To perform this normalization, the forward direction of the head (headset) is projected onto the horizontal plane. The transformation applied to all tracked objects (left hand controller, right hand controller) is the rotation about the vertical axis so that the projected forward direction of the head aligns with the forward direction of the coordinate system.

% \textcolor{red}{TODO: cut down to just describe the features in a paragraph or so. 17 DOF, technically reduced without head yaw.}
At each time step (frame) processed by the VR device, the position and orientation of the user's left hand, right hand, and head are captured. Three positional coordinates and four orientation coordinates (in quaternion format) are captured for each of the three tracked objects, totaling 21 dimensions captured per frame. After the body-relative transformation is applied, 18 dimensions remain, as the three positional coordinates of the head are eliminated by this transformation. We also lose one rotational degree of freedom (yaw) but the quaternion representation still encodes the remaining rotation with all four values. The values of interest to us are the first and second derivatives of these 18 values; the result is 36 values per frame describing \textit{body-relative velocity} and \textit{body-relative acceleration}.

Each user's VR device may render frames at a slightly different frequency due to a variety of external factors. To eliminate frame rate as a potential confounding factor, we first normalize all motion capture streams to a constant 30 frames per second by using a numerical linear interpolation for positional coordinates and a spherical linear interpolation for orientation quaternions. Each session of a user was then split into 30-second sequences. The selection of 30 seconds was due to better performance than with the 1-second blocks used in previous work, perhaps due to the shift from static (position) to dynamic (velocity/acceleration) features. Future investigation of this parameter would be beneficial.

In summary, an individual sequence has 30 seconds, 30 frames a second, and 36 values per frame; thus, our model has an input shape of $(900 \times 36)$ consisting of both velocity and acceleration characteristics.

\subsection{Model}
\label{sec:architecture}

The model's task is to identify a user based upon their motion. More formally, the model is given a $(900 \times 36)$ sequence as described above. With that sequence, the model ought to predict the participant who generated that motion, represented as a value of a categorical variable encoded with a one-hot encoding.

The model we have selected is a Long Short-Term Memory (LSTM) model \cite{hochreiter1997long}, implemented in Python version 3.10.2 using Keras version 2.10.1.
The choice of LSTM was to take advantage of the sequential nature of the data. Most hyperparameters for the model were left to the defaults; in particular, the Adam optimizer \cite{kingma2014adam} was used with a learning rate of $0.001$. Specifically, we utilize the ``LSTM Funnel'' architecture described by Nair and collaborators \cite{nair2023deep}.

The predictions were made per session by taking the entire session
of pose tracking data, computing 30-second sequences as described above, and then summing the logarithmic probability of each user reported by the model across all samples. We interpreted this distribution as a probability estimation for the classification of the session
as a whole, in line with previous work \cite{Miller2020}.

\subsection{Evaluation}
\label{ssec:eval}

We report three metrics for evaluation. Identification-focused works \cite{Pfeuffer2019,Miller2020,Moore2021,Miller2022}
almost exclusively use accuracy for the model's evaluation metric. However, accuracy varies significantly as the number of classes varies, both in theory (as both false positive and negative identification rate depends on the number of potential identities to match against) and in practice (see \cite{Pfeuffer2019,Miller2020,Wang2021}). This is further described in \cite{miller_comparing_2024}. To address this issue, we seek an evaluation metric that is invariant to the number of classes to predict upon. One such metric is \textit{multiclass
AUC}, defined by Hand and Till \cite{Hand2001}. In short, multiclass AUC can
be described as the likelihood a randomly chosen sample will be identified as its true class as opposed to a randomly chosen other class. In order to compare against previous work, which does not use multiclass AUC, we use $N$-class accuracy, which is an estimate of the expected accuracy of the model if it had been tested on only $N$ classes.

\section{Results}

\label{sec:results}

The focus of this work is on the effect of duration and delay on accuracy. While these topics have been explored in previous work \cite{Miller2022, Moore2021}, they have not been given names.

First is \textit{duration}. We use this word to refer to the length of time of a set of data covers. This is relevant to both training (enrollment) and testing (query) periods, so one can speak of the training duration and the testing duration separately. 
    
Second is \textit{delay}. This is the amount of physical time in between the data representing the training and the testing portions of the data. For example, taking a 15 minute recording and training on the first 10 minutes and testing on the last 5 minutes would have minimal delay \cite{Miller2020}. On the other hand, collecting data over the course of a week and then asking participants to return nearly a year later \cite{Miller2022} would have a very high delay.

\subsection{Identification by Delay and Duration}

First, we demonstrate that the motion data in question is effective at identification, and demonstrate that the selection of duration and delay can greatly affect the identifiability of an activity. The delay is here translated into the train-test split, which is either between or within sessions. A split \textit{between} sessions indicates that if data from one weekly session (that is, a recording of a person in a given week) is used as training data, then testing data cannot be drawn from the same session. In particular, we use the first six weeks from all participants for training, and the last two weeks for testing. A split \textit{within} session allows training and testing data to both be drawn from the same weekly session. In particular, the training data was 4/5ths of each session duration, with at least two minutes of buffer time between the train and test sections. The test duration is simply the duration of data which is tested, described in the table. For comparison, this was on average about 5 minutes per session. Similar to previous work \cite{Miller2020,Nair2022}, predictions are made using a sliding window of thirty seconds that has a step size of one-second intervals, determining a prediction for each of these segments, and then aggregating these into a single prediction by selecting the most commonly predicted identity. Results are given in Table \ref{tab:ident} that show each accuracy
metric for the 183 participants who had at least 5 unique sessions and 2 hours of data total.

\begin{table}[h]
\caption{Evaluation of identification models by train-test split method (between or within) and test size.}
\label{tab:ident} %
\resizebox{\linewidth}{!}{%
\begin{tabular}{ccllll}
\toprule 
Split & Test Duration & Accuracy  & Multiclass AUC  & 30-Class Accuracy\tabularnewline
\midrule 
Between & $\sim$5 min & 49.18\%  & 93.40\%  & 68.19\% \tabularnewline
Between & 2 sessions $\times$ $\sim$25 minutes & 77.60\%  & 98.31\%  & 87.09\% \tabularnewline
Within & $\sim$5 min & 71.70\%  & 98.71\% & 87.00\% \tabularnewline
Within & 8 sessions $\times$ $\sim$5 minutes & 100.00\%  & 100.00\%  & 100.00\% \tabularnewline
\bottomrule
\end{tabular}
}
\end{table}

The results indicate that this pipeline is effective for identifying this kind of motion data and that both test duration and delay influence accuracy. A longer duration implies more data, and more data almost always leads to better predictions. A greater delay leads to worse identifiability, as more aspects of the participant’s behavior can change over that greater span of time. Other works have shown a similar trend \cite{Moore2021,Miller2022}. 

What is worth noting, though, is how dramatically the accuracy values can change. In the between-session case, there is 50\% accuracy on the full 183-person dataset. Meanwhile, by increasing the test duration and decreasing the delay, accuracy rises to 100\% upon the same dataset.

\subsection{Identification by Delay\label{subsec:Delay}}

To study delay, we vary the weeks
upon which a model is trained and tested while keeping the duration
the same. Because the focus is on the upper end of delay, we do not look at the within-session splits, but instead focus entirely on between-session splits.

\begin{figure}[h]
\centering \includegraphics[width=3.3in]{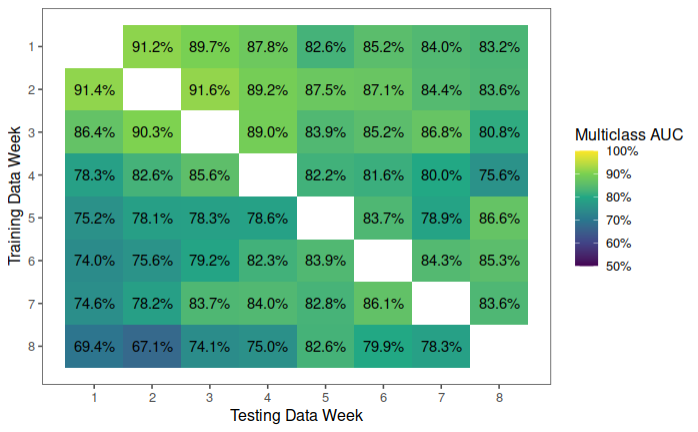} \caption{Separating the training and testing sets by larger time reduces accuracy.
The x-axis and y-axis are the testing and training weeks, respectively.
The panels are colored indicating identifiability (operationalized as multiclass AUC), with yellow as a higher
accuracy. Note a trend that higher multiclass AUC is along the diagonal
(i.e., minimal delay).}
\label{fig:timing_delay} 
\end{figure}

The multiclass AUCs reported
in Figure \ref{fig:timing_delay} are produced by training the model upon one week's worth of data and testing it on a different
week's worth of data. In total, there are $8\times7=56$ entries. All data for the selected training
session is used, and all testing data matching a participant in the
training set is tested with. Note that multiclass AUC is reported
both for pairs where training week happens before testing week, as
would be expected for an attacker, but also in pairs where testing
week happens before training week, which is relevant to pose re-identification
well after data collection.

The results in Figure \ref{fig:timing_delay} show a pattern that
multiclass AUC is higher when training and testing sessions have less
delay (i.e, at the near-diagonals, especially in the top left) than when there is more delay (near
bottom left and top right). This effect varies somewhat across weeks.

To confirm the statistical significance of this effect over weeks, a mixed-effect model was fit using the software package ``lmerTest'' to the 56 data points of multiclass AUC shown in Figure \ref{fig:timing_delay}. This model fit the logit values of multiclass AUC based upon delay (the positive integer difference in number of weeks between training and testing) with random intercepts for training week. A random intercept for testing week was included, but it resulted in a singular fit, accounting for no variance, and was dropped from the model. The effect of delay upon multiclass AUC is highly significant ($t(45.57)=-9.752$, $p=6.3\times10^{-13}$), with an intercept of 1.96 logits (87.74\%) and a slope of -0.12 logits per additional week of delay. These results estimate a one-week delay to have a multiclass AUC of 86.30\% and a seven-week delay to have a multiclass AUC of 74.44\%.

While this is evidence that identification decreases over time, it is a small decrease. With a larger dataset like BOXRR \cite{nair2023berkeley} upon which a model can attain a lower bound multiclass AUC of 0.999975 \cite{nair_unique_2023}, and assuming that the identification rate decreases at the same slope of -0.12 logits per week, one could estimate it would take over 40 weeks to drop to 90\% accuracy within a set of 10. While extrapolation should be done with caution, it is clear that identification by motion, even over a long duration, is plausible.

\subsection{Identification by Duration\label{subsec:Duration}}

To study duration, we vary
the number of separate sessions in the training set and the training
time per session to investigate its effect on multiclass AUC. 

\begin{figure}[h]
\centering \includegraphics[width=3.3in]{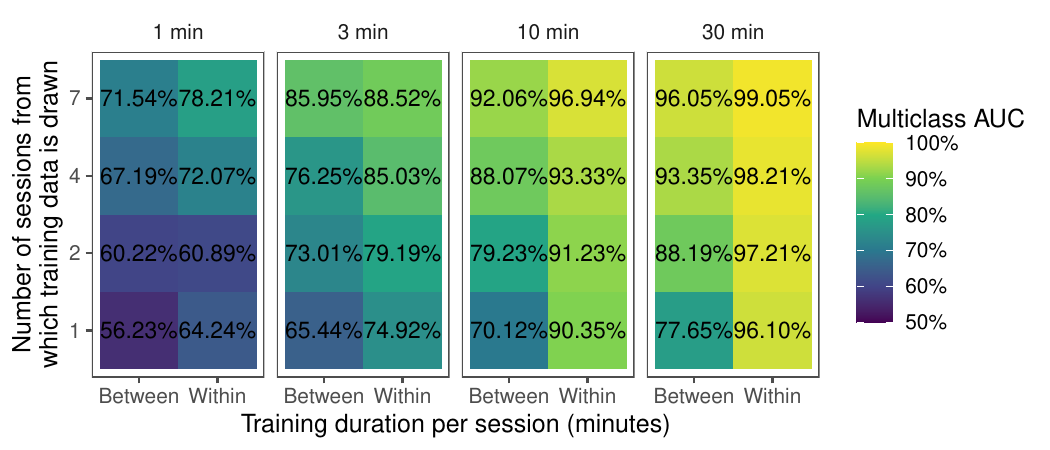} 
\caption{Number of sessions and duration of each session affect identifiability, operationalized as multiclass AUC. Two
panels shown horizontally indicate whether the comparison is drawn
between sessions or within the same session. Within each panel, the x-axis indicates the training duration per session, and y-axis indicates the number of sessions.
The rectangles are colored indicating identifiability, with yellow as a higher
accuracy.}
\label{fig:timing_duration} 
\end{figure}

In the first analysis, the train-test split was performed by first
randomly selecting a set of training sessions of size at most 1, 2,
4, or 7 for each participant, leaving at least one session
for testing. For example, in the case seven sessions were requested
but a participant only took part in six, five of those six were used
for training and one was held out for between-sessions testing. Of
the selected training sessions, spans of time for training and within-sessions
testing were chosen. Note that this does not ensure perfectly equivalent delays because a selection of more training data is more likely to be nearer in time to the test data. This is a limitation. Additionally, due to limited spans of data available, the average training span for each of the 1, 3,
10, and 30 minute conditions were durations of 1:00, 2:59, 9:39, and
22:32 respectively. Sessions shorter than eight minutes total were
dropped from this analysis. Results are given
in Figure \ref{fig:timing_duration}.

In both panels, it is visible that an increase in duration leads to an increase in accuracy. Increasing the training data duration per session (1 minute to 30 minutes) and the number of sessions (1 to 7) both produce significant gains in accuracy. 

There is only one exception, which is when accuracy decreases when adding a second session to the 1 minute within-session test. We hypothesize this is because the characteristics upon which one session can be identified appear to be different from the characteristics across sessions.

While this analysis is focused on duration, there is also a finding on delay. With session number and duration per session held constant, in every case, the within-session AUC was greater than the between-session AUC. The differences are most stark when fewer sessions are used. For example, when up to 30 minutes of one session per participant is used for training, the model achieves an AUC of 77.65\% with the between-session data and a 96.10\% with the within-session data. This is the difference between attaining 10.93\% rank-1 accuracy (20 of 183) and 43.17\% rank-1 accuracy (79 of 183). 

This demonstrates that duration matters significantly, and using these techniques, it takes only minutes of motion to identify someone with fair accuracy.

\section{Discussion}
% TODO: better summary here.

\subsection{Summary of Results}

We investigate the effects of delay and duration upon identifiability and find that an increased delay between training and testing recordings decreases accuracy, and an increased duration of training data increases accuracy. Overall, given that human motion is a complex process with many components and interactions, we infer that some of these factors may be consistent on short time scales and some on long time scales. Future work
ought not to look at one time scale but many.

% milticlass auc
In response to previous work with varying identification sizes,
we select and justify the Multiclass AUC evaluation metric to evaluate
identifiability across sample sizes. Removing
this confound can let future work clarify other important trends
in accuracy, such as time, feature selection, or activity.

\subsection{Implications for Privacy}
% TODO: make this a lot better.

This work continues to survey the risks that VR poses to privacy.
The most important question in this space is how identifying various
data sources, situations, and activities are, what makes these identifying,
and what can be done about it. By understanding what influences the
accuracy of de-anonymization techniques, researchers can develop more
effective and more efficient ways to limit risk to end users.

We encourage future researchers to continue to investigate the effect of delay on identifiability in their own datasets. This includes focusing on between-session identification, as is also highlighted by \cite{Moore2021}. Within-session identification can lead to unrealistically high accuracies. Second, we encourage other researchers to report not simply accuracy but also multiclass AUC so that model performance can be adequately compared across classification sizes.

\subsection{Limitations and Future Work}

Some limitations of this work related to the dataset under study, the SLVRClaD dataset. These include that while participants knew
their pose tracking data was collected, they were not aware what features
of their data would be most identifying so that they could change
their behavior to avoid being tracked, e.g. vary their height week-to-week
to fool the model. All participants used the same headset for the
entire duration of the study, which according to previous work \cite{miller2020within,Miller2021}
can make identification easier.

Regarding attack models, some avenues for future work include demonstrating
effective attacks beyond biometrics. For example, depending on what
is transmitted, almost all of a target's visual and auditory experience
can be recorded or inferred. This includes inferences
about the target's attention to objects, content, or people due to both conscious and unconscious mechanisms. Future work can explore the potential of varying the 30s segment size as well as determining which signals are stable or temporary.

\section{Conclusion}

This research continues to probe the privacy risks associated with the collection and transmission of headset and hand controller motion in consumer virtual reality (VR) devices. The findings underscore the robustness of identifiability in VR-tracked motion data, even with varying signal degradations. We emphasize the need for heightened consumer awareness and the development of defenses \cite{nair2023deep} against re-identification in scenarios where anonymity is desired. As social VR gains popularity, the potential privacy risks within the metaverse become increasingly apparent. 

\section*{Acknowledgment}

This work was supported in part by the Sozo Graduate Fellowship, Minderoo Foundation, National Science Foundation, National Physical Science Consortium, Fannie and John Hertz Foundation, and Berkeley Center for Responsible Decentralized Intelligence.

\bibliographystyle{IEEEtran}
\bibliography{biblio}

\begin{thebibliography}{10}
\providecommand{\url}[1]{#1}
\csname url@samestyle\endcsname
\providecommand{\newblock}{\relax}
\providecommand{\bibinfo}[2]{#2}
\providecommand{\BIBentrySTDinterwordspacing}{\spaceskip=0pt\relax}
\providecommand{\BIBentryALTinterwordstretchfactor}{4}
\providecommand{\BIBentryALTinterwordspacing}{\spaceskip=\fontdimen2\font plus
\BIBentryALTinterwordstretchfactor\fontdimen3\font minus \fontdimen4\font\relax}
\providecommand{\BIBforeignlanguage}[2]{{%
\expandafter\ifx\csname l@#1\endcsname\relax
\typeout{** WARNING: IEEEtran.bst: No hyphenation pattern has been}%
\typeout{** loaded for the language `#1'. Using the pattern for}%
\typeout{** the default language instead.}%
\else
\language=\csname l@#1\endcsname
\fi
#2}}
\providecommand{\BIBdecl}{\relax}
\BIBdecl

\bibitem{Pfeuffer2019}
\BIBentryALTinterwordspacing
K.~Pfeuffer, M.~J. Geiger, S.~Prange, L.~Mecke, D.~Buschek, and F.~Alt, ``{Behavioural Biometrics in VR: Identifying People from Body Motion and Relations in Virtual Reality},'' in \emph{Proceedings of the 2019 CHI Conference on Human Factors in Computing Systems}, ser. CHI '19.\hskip 1em plus 0.5em minus 0.4em\relax New York, NY, USA: ACM, 2019, pp. 110:1----110:12. [Online]. Available: \url{http://doi.acm.org/10.1145/3290605.3300340}
\BIBentrySTDinterwordspacing

\bibitem{Moore2021}
A.~G. Moore, R.~P. McMahan, H.~Dong, and N.~Ruozzi, ``{Personal identifiability and obfuscation of user tracking data from VR training sessions},'' \emph{Proceedings - 2021 IEEE International Symposium on Mixed and Augmented Reality, ISMAR 2021}, pp. 221--228, 2021.

\bibitem{Miller2020}
\BIBentryALTinterwordspacing
M.~R. Miller, F.~Herrera, H.~Jun, J.~A. Landay, and J.~N. Bailenson, ``{Personal identifiability of user tracking data during observation of 360-degree VR video},'' \emph{Scientific Reports}, vol.~10, no.~1, pp. 17\,404--17\,413, 2020. [Online]. Available: \url{https://doi.org/10.1038/s41598-020-74486-y}
\BIBentrySTDinterwordspacing

\bibitem{Nair2022}
\BIBentryALTinterwordspacing
V.~Nair, G.~M. Garrido, D.~Song, and J.~F. O'Brien, ``Exploring the privacy risks of adversarial {VR} game design,'' \emph{Proc. Priv. Enhancing Technol.}, vol. 2023, no.~4, pp. 238--256, 2023. [Online]. Available: \url{https://doi.org/10.56553/popets-2023-0108}
\BIBentrySTDinterwordspacing

\bibitem{Olade2020}
I.~Olade, C.~Fleming, and H.~N. Liang, ``{Biomove: Biometric user identification from human kinesiological movements for virtual reality systems},'' \emph{Sensors (Switzerland)}, vol.~20, no.~10, pp. 1--19, 2020.

\bibitem{mathis_knowledge-driven_2020}
\BIBentryALTinterwordspacing
F.~Mathis, H.~I. Fawaz, and M.~Khamis, ``\BIBforeignlanguage{en}{Knowledge-driven {Biometric} {Authentication} in {Virtual} {Reality}},'' in \emph{\BIBforeignlanguage{en}{Extended {Abstracts} of the 2020 {CHI} {Conference} on {Human} {Factors} in {Computing} {Systems}}}.\hskip 1em plus 0.5em minus 0.4em\relax Honolulu HI USA: ACM, Apr. 2020, pp. 1--10. [Online]. Available: \url{https://dl.acm.org/doi/10.1145/3334480.3382799}
\BIBentrySTDinterwordspacing

\bibitem{Wang2021}
X.~Wang and Y.~Zhang, ``{Nod to Auth: Fluent AR/VR Authentication with User Head-Neck Modeling},'' \emph{Conference on Human Factors in Computing Systems - Proceedings}, 2021.

\bibitem{liebers_using_2021}
\BIBentryALTinterwordspacing
J.~Liebers, P.~Horn, C.~Burschik, U.~Gruenefeld, and S.~Schneegass, ``\BIBforeignlanguage{en}{Using {Gaze} {Behavior} and {Head} {Orientation} for {Implicit} {Identification} in {Virtual} {Reality}},'' in \emph{\BIBforeignlanguage{en}{Proceedings of the 27th {ACM} {Symposium} on {Virtual} {Reality} {Software} and {Technology}}}.\hskip 1em plus 0.5em minus 0.4em\relax Osaka Japan: ACM, Dec. 2021, pp. 1--9. [Online]. Available: \url{https://dl.acm.org/doi/10.1145/3489849.3489880}
\BIBentrySTDinterwordspacing

\bibitem{Falk2021}
B.~Falk, Y.~Meng, Y.~Zhan, and H.~Zhu, ``{POSTER: ReAvatar: Virtual Reality De-anonymization Attack through Correlating Movement Signatures},'' \emph{Proceedings of the ACM Conference on Computer and Communications Security}, pp. 2405--2407, 2021.

\bibitem{Miller2021}
R.~Miller, N.~K. Banerjee, and S.~Banerjee, ``{Using siamese neural networks to perform cross-system behavioral authentication in virtual reality},'' \emph{Proceedings - 2021 IEEE Conference on Virtual Reality and 3D User Interfaces, VR 2021}, pp. 140--149, 2021.

\bibitem{Miller2022}
------, ``{Temporal Effects in Motion Behavior for Virtual Reality (VR) Biometrics},'' \emph{Proceedings - 2022 IEEE Conference on Virtual Reality and 3D User Interfaces, VR 2022}, pp. 563--572, 2022.

\bibitem{miller2020within}
------, ``Within-system and cross-system behavior-based biometric authentication in virtual reality,'' in \emph{2020 IEEE Conference on Virtual Reality and 3D User Interfaces Abstracts and Workshops (VRW)}.\hskip 1em plus 0.5em minus 0.4em\relax IEEE, 2020, pp. 311--316.

\bibitem{Shen2018}
Y.~Shen, H.~Wen, C.~Luo, W.~Xu, T.~Zhang, W.~Hu, and D.~Rus, ``{GaitLock: Protect Virtual and Augmented Reality Headsets Using Gait},'' \emph{IEEE Transactions on Dependable and Secure Computing}, vol. 5971, no.~c, pp. 1--14, 2018.

\bibitem{schell_extensible_2023}
\BIBentryALTinterwordspacing
C.~Rack, K.~Kobs, T.~Fernando, A.~Hotho, and M.~E. Latoschik, ``Extensible {Motion}-based {Identification} of {XR} {Users} with {Non}-{Specific} {Motion},'' Feb. 2023, arXiv:2302.07517 [cs]. [Online]. Available: \url{http://arxiv.org/abs/2302.07517}
\BIBentrySTDinterwordspacing

\bibitem{sivasamy_vrcauth_2020}
M.~Sivasamy, V.~N. Sastry, and N.~P. Gopalan, ``{VrCauth}: {Continuous} authentication of users in virtual reality environment using head-movement,'' \emph{Proceedings of the 5th International Conference on Communication and Electronics Systems, ICCES 2020}, no. Icces, pp. 518--523, 2020, iSBN: 9781728153711.

\bibitem{wierzbowski_behavioural_2022}
\BIBentryALTinterwordspacing
M.~Wierzbowski, G.~Pochwatko, P.~Borkiewicz, D.~Cnotkowski, M.~Pabis-Orzeszyna, and P.~Kobylinski, ``\BIBforeignlanguage{en}{Behavioural {Biometrics} in {Virtual} {Reality}: {To} {What} {Extent} {Can} {We} {Identify} a {Person} {Based} {Solely} on {How} {They} {Watch} 360-{Degree} {Videos}?}'' in \emph{\BIBforeignlanguage{en}{2022 {IEEE} {International} {Symposium} on {Mixed} and {Augmented} {Reality} {Adjunct} ({ISMAR}-{Adjunct})}}.\hskip 1em plus 0.5em minus 0.4em\relax Singapore, Singapore: IEEE, Oct. 2022, pp. 417--422. [Online]. Available: \url{https://ieeexplore.ieee.org/document/9974570/}
\BIBentrySTDinterwordspacing

\bibitem{mustafa_unsure_2018}
T.~Mustafa, R.~Matovu, A.~Serwadda, and N.~Muirhead, ``Unsure {How} to {Authenticate} on {Your} {VR} {Headset}?'' in \emph{{IWSPA}'18: 4th {ACM} {International} {Workshop} on {Security} {And} {Privacy} {Analytics}.}\hskip 1em plus 0.5em minus 0.4em\relax New York, NY, USA: ACM, 2018, pp. 23--30.

\bibitem{liebers_exploring_2023}
\BIBentryALTinterwordspacing
J.~Liebers, C.~Burschik, U.~Gruenefeld, and S.~Schneegass, ``\BIBforeignlanguage{en}{Exploring the {Stability} of {Behavioral} {Biometrics} in {Virtual} {Reality} in a {Remote} {Field} {Study}},'' in \emph{\BIBforeignlanguage{en}{29th {ACM} {Symposium} on {Virtual} {Reality} {Software} and {Technology}}}.\hskip 1em plus 0.5em minus 0.4em\relax Christchurch New Zealand: ACM, Oct. 2023, pp. 1--12. [Online]. Available: \url{https://dl.acm.org/doi/10.1145/3611659.3615696}
\BIBentrySTDinterwordspacing

\bibitem{nair_unique_2023}
\BIBentryALTinterwordspacing
V.~Nair, W.~Guo, J.~Mattern, R.~Wang, J.~F. O'Brien, L.~Rosenberg, and D.~Song, ``Unique {Identification} of 50,000+ {Virtual} {Reality} {Users} from {Head} \& {Hand} {Motion} {Data},'' Feb. 2023, arXiv:2302.08927 [cs]. [Online]. Available: \url{http://arxiv.org/abs/2302.08927}
\BIBentrySTDinterwordspacing

\bibitem{Liebers2021}
J.~Liebers, M.~Abdelaziz, L.~Mecke, A.~Saad, J.~Auda, U.~Grunefeld, F.~Alt, and S.~Schneegass, ``Understanding user identification in virtual reality through behavioral biometrics and the efect of body normalization,'' \emph{Conference on Human Factors in Computing Systems - Proceedings}, 2021, {ISBN}: 9781450380966.

\bibitem{garrido2023sok}
G.~M. Garrido, V.~Nair, and D.~Song, ``Sok: Data privacy in virtual reality,'' 2023.

\bibitem{Han2023}
\BIBentryALTinterwordspacing
E.~Han, M.~R. Miller, C.~DeVeaux, H.~Jun, K.~L. Nowak, J.~T. Hancock, N.~Ram, and J.~N. Bailenson, ``{People, places, and time: a large-scale, longitudinal study of transformed avatars and environmental context in group interaction in the metaverse},'' \emph{Journal of Computer-Mediated Communication}, vol.~28, no.~2, 01 2023, zmac031. [Online]. Available: \url{https://doi.org/10.1093/jcmc/zmac031}
\BIBentrySTDinterwordspacing

\bibitem{rack_who_2023}
\BIBentryALTinterwordspacing
C.~Rack, T.~Fernando, M.~Yalcin, A.~Hotho, and M.~E. Latoschik, ``Who {Is} {Alyx}? {A} new {Behavioral} {Biometric} {Dataset} for {User} {Identification} in {XR},'' Aug. 2023, arXiv:2308.03788 [cs]. [Online]. Available: \url{http://arxiv.org/abs/2308.03788}
\BIBentrySTDinterwordspacing

\bibitem{hochreiter1997long}
S.~Hochreiter and J.~Schmidhuber, ``Long short-term memory,'' \emph{Neural computation}, vol.~9, no.~8, pp. 1735--1780, 1997.

\bibitem{kingma2014adam}
D.~P. Kingma and J.~Ba, ``Adam: A method for stochastic optimization,'' \emph{arXiv preprint arXiv:1412.6980}, 2014.

\bibitem{nair2023deep}
V.~Nair, W.~Guo, J.~F. O'Brien, L.~Rosenberg, and D.~Song, ``Deep motion masking for secure, usable, and scalable real-time anonymization of virtual reality motion data,'' 2023.

\bibitem{miller_comparing_2024}
M.~R. Miller, ``Comparing identification accuracy across varying classification set sizes with multiclass {AUC},'' Feb. 2024.

\bibitem{Hand2001}
D.~J. Hand and R.~J. Till, ``{A Simple Generalisation of the Area Under the ROC Curve for Multiple Class Classification Problems},'' \emph{Machine Learning}, vol.~45, no.~2, pp. 171--186, 2001.

\bibitem{nair2023berkeley}
V.~Nair, W.~Guo, R.~Wang, J.~F. O'Brien, L.~Rosenberg, and D.~Song, ``Berkeley open extended reality recordings 2023 (boxrr-23): 4.7 million motion capture recordings from 105,852 extended reality device users,'' 2023.

\end{thebibliography}

\end{document}